# Achieving Selective Damage Interrogation and Sub-Wavelength Resolution in Thin Plates with Embedded Metamaterial Acoustic Lenses

F. Semperlotti[*], H. Zhu

*Department of Aerospace and Mechanical Engineering, University of Notre Dame, Notre Dame, Indiana 46556, USA*

In this study, we present an approach to ultrasonic beamforming and high resolution identification of acoustic sources having critical implications for structural health monitoring technology. The proposed concept is based on the design of dynamically tailored structural elements via embedded acoustic metamaterial lenses. This approach provides a completely new alternative to conventional phased-array technology enabling the formation of steerable and collimated (or focused) ultrasonic beams by exploiting a single transducer. The ultrasonic beams can be steered by simply tuning the frequency of the excitation. Also, the embedded lens can be designed to achieve sub-wavelength resolution to incipient clustered damage.

[*]Corresponding author e-mail fsemperl@nd.edu

Selective interrogation is considered as a critical enabling technology for the implementation of the next generation of ultrasonic based Structural Health Monitoring (SHM) systems. The ability to send ultrasonic energy in a preferential direction leads to increased damage sensitivity due to improved interaction (either in terms of back-scattered echo for a linear damage or of the nonlinear harmonic amplitude for nonlinear incipient damage[1,2]) between the interrogation signal and the damage. In highly directional or anisotropic material, such as for layered composite structures, the direction of energy propagation can be largely different from the original direction of the interrogation signal[3]. This situation results in reduced damage sensitivity because only a fraction of the incident wave energy can effectively reach the damage. The ability to generate highly directional and collimated signals can be exploited to compensate for this intrinsic characteristic of the material. In case of a multiple damage scenario, a directional interrogation would also allow to selectively scan the structural element and acquire data from the individual damage, which will increase the sensitivity and provide additional information for damage localization.

To-date, one of the most diffused approach to achieve selective interrogation for SHM applications has certainly been based on Phased-Arrays (PA) technology[4,5]. PA exploits a set of transducers activated according to pre-defined time delays in order to produce either directional wavefronts or focused excitation at a prescribed spatial location. Although a robust and, to some extent, effective technology PA exhibits two important limitations that prevent its extensive use in practical applications. The first limitation consists in the large number of transducers required for implementation. The need for an extended transducer network is regarded as a major limitation in SHM applications because strictly related to increased probability of false alarms and hardware malfunctions as well as higher system complexity that affects fabrication and installation (e.g. harnessing, powering, etc.). The second major drawback of PA technology is related to its inability to generate collimated signals. In PAs, either the directional or focused excitation is the result of constructive interference produced by the superposition of multiple omni-directional wavefronts. In a multiple damage scenario, these wavefronts produce multiple reflected echoes (although weaker than those generated at the focal point) that reduce the accuracy of the detection. The ability to create directional, collimated ultrasonic beams would greatly benefit the damage detection and localization process by enabling a selective scan of prescribed structural areas. In addition, when multiple damages are closely spaced together (i.e. clustered damage) the damage signature does not provide the level of spatial resolution necessary to discern the individual damage. This situation typically results in an overestimated damage footprint and in lack of information about the damage shape.

In this Letter, we present an approach targeted to address these issues during both the interrogation and sensing phase. In particular, we develop a selective interrogation technique that overcomes the limitations of PAs by producing highly collimated (or focused), steerable ultrasonic beams by using a single transducer. We also show how the same technology can be used to increase the sensitivity to clustered damage by enabling sub-wavelength resolution.

The proposed approach relies on the concept of dynamic structural tailoring of the host structure achieved via acoustic metamaterial based design. By exploiting the characteristic behavior of anisotropic resonant metamaterials[6,7], acoustic lenses can be designed and embedded (or surface mounted) into structural elements to mold the ultrasonic wavefronts generated by either a single transducer (*actuation*) or by a nonlinear damage (*sensing*). Anisotropic resonant acoustic metamaterials in a fluid background[8] were proven to exhibit collimation and sub-wavelength resolution capabilities due to the characteristic hyperbolic nature of their Equi-Frequency-Contours (EFC)[9-11]. We exploit this property in conjunction with spatially tailored frequency bandgaps[12,13] and localized modes[14,15] to design embedded acoustic lenses for selective interrogation. The operating principle is illustrated on bulk materials and then extended to a finite structure for practical application. We consider a bulk perfectly periodic elastic metamaterial with a square lattice structure (lattice constant $a$) made of an aluminum background and cylindrical steel inclusions ($r_s=0.25a$) coated in silicon rubber ($r_e=0.35a$). The dispersion curves along the first Brillouin zone (FIG. 1a), calculated by Plane Wave Expansion[16,17] (PWE), show the existence of a full bandgap for the SV mode in the non-dimensional frequency range $\Omega = \omega a/2\pi c_t = 0.27 \div 0.4$. In this range, propagating waves are not supported therefore the material effectively acts as a mechanical stopband filter. The response of the bulk material in the bandgap can be tailored by exploiting defects that locally break the periodicity of the crystal[17]. Defects are associated with a spatially localized dynamic response which results in additional dispersion curves located inside the bandgap. Among the different types of defects, line defects (also called waveguides) can be used to create spatially localized modes inside the bandgap[17]. These modes are confined by the defect but are free to propagate inside it. Defects can be created by altering either the geometric or the material properties of one or more inclusions. For the material under study, a line defect was created by altering the size of the steel inclusions ($r_s=0.15a$) in the center column of a squared lattice material (see FIG. 2a, inset). The introduction of a line defect results in additional localized modes inside the bandgap (red curves in FIG. 1b) that propagate though the crystal in the direction of the defect, as shown by the modal displacement field in FIG. 1b. These localized modes result in preferential paths of propagation through the crystal at specific frequencies.

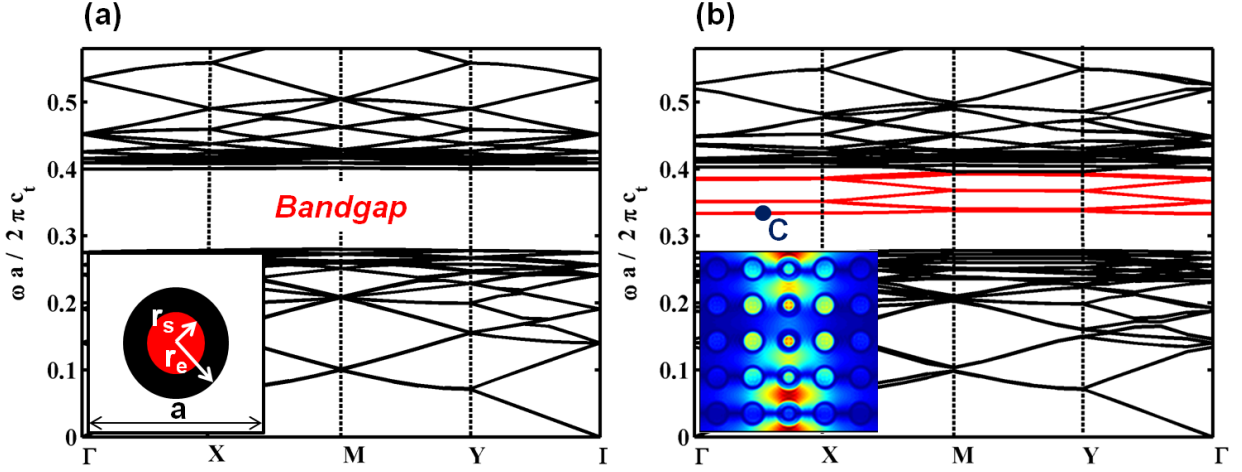

FIG. 1. Dispersion characteristics (SV mode) in the first Brillouin zone of (a) the perfectly periodic and (b) defected material showing the existence of a full bandgap and the generation of localized modes (red lines) inside the bandgap due to the defect. The inset in (a) shows the geometric parameters of the main unit cell while the inset in (b) shows the displacement field of the localized defect mode corresponding to point C.

The frequency of the localized modes can be controlled by properly designing the properties of the defect. It follows that the propagation characteristics of a metamaterial can be tailored in both the spatial and frequency domain by designing a network of defects.

The second key characteristic exploited in this approach is related to the ability of anisotropic resonant metamaterials of generating hyperbolic Equi-Frequency-Contours (EFC). The direct comparison of the EFC curves for the SV mode (FIG. 2) of a square lattice $a \times a$ resonant metamaterial (isotropic in the long wavelength limit) with a rectangular lattice $a \times 4a$ resonant metamaterial (anisotropic in the long wavelength limit) shows that the anisotropic nature of the rectangular lattice results in the generation of hyperbolic EFCs[8].

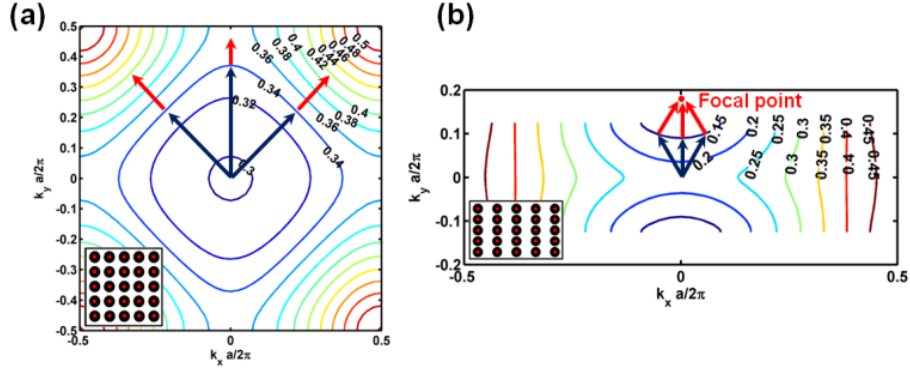

FIG. 2. Comparison of the Equi-Frequency-Contours for the (a) isotropic and (b) anisotropic resonant metamaterial. The superimposed schematic shows that for the hyperbolic EFCs (b) the initial direction of propagation of the incident wave (indicated by the wavevector, blue arrows) is steered to become orthogonal to the EFC branches (that is the direction of the group velocity, red arrows). Therefore, the initially diffused field is focused at the focal point of the hyperbola as it travels through the crystal.

Depending on the specific design of the hyperbolic EFCs and on the location of the focal point, the metamaterial will act either as a focusing or a collimating lens for incident ultrasonic waves. In fact, an incident wave initially emitted in the direction indicated by the wavevector (FIG. 2, blu arrows) is redirected so that in the far-field the group velocity becomes orthogonal to the EFCs (FIG. 2, red arrows). Therefore, an initially diffused field propagating through an anisotropic crystal with hyperbolic EFCs will be converted into a focused wave field while travelling through the crystal. The location of the focal point depends on the specific design of the lattice structure that ultimately affects the shape of the hyperbolic EFCs. In the limit case of a flat EFC, the focal point of the hyperbola moves to infinity therefore originating, in the far-field, a perfectly collimated wave field.

These characteristics of anisotropic resonant metamaterials can be exploited in the design of embedded lenses for SHM applications. In particular, we show that anisotropic metamaterial lenses with spatially tailored line defects can be used to create focused (or collimated) and steerable excitation by using a single ultrasonic transducer. This concept is illustrated using a 2m×2m simply supported thin aluminum plate with a thickness $t$=3 mm (FIG. 3). The inclusions are represented by mass-in-mass systems. The overall structure is modeled according to the Kirchhoff's thin plate theory with attached lumped single degree of freedom resonators. The governing equation of the plate is discretized using the assumed modes method[18] and solved by direct time integration. The resulting system of governing equations is given by:

$$M_s^{pq}\ddot{q}_s(t) + C_s^{pq}\dot{q}_s(t) + K_s^{pq}q_s(t) + k_R\left\{\sum_{pq}\phi_{pq}(x_R^r, y_R^r)q_s(t) - q_R(t)\right\}\phi_{pq} = \phi_{pq}(x_R, y_R)F(t) \quad (1)$$

$$m_R^r\ddot{q}_R^r(t) + k_R^r\left\{q_R^r(t) - \sum_{pq}\phi_{pq}(x_R^r, y_R^r)q_s(t)\right\} = 0 \quad (2)$$

with $p$=1…$n$, $q$=1…$m$, and $r$=1…$L$. Eqns. (1) and (2) represent a $n \times m \times L$ system of second order coupled ordinary differential equations where $M_s^{pq}$, $C_s^{pq}$ and $K_s^{pq}$ are generalized mass, damping and stiffness matrices of the plate associated with the $pth \times qth$ assumed modes, $q_s(t)$ and $q_R(t)$ are the generalized coordinates of the plate and of the local resonators, $m_R^r$ and $k_R^r$ are the mass and stiffness of the resonator, $\phi_{pq}$ is the set of basis functions used for the discretization.

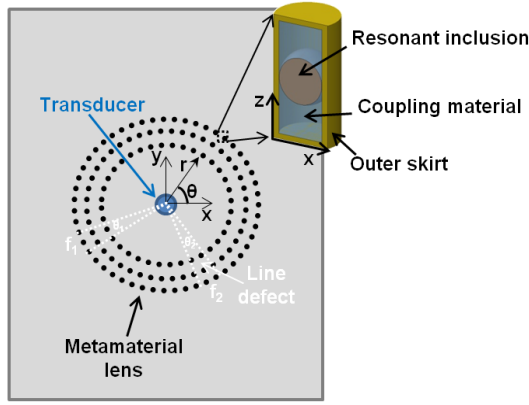

FIG. 3. Schematic of the thin aluminum plate with the embedded metamaterial lens. Selected sectors of the lens at prescribed azimuthal locations $\theta_i$ can be tailored (using line defects) to create localized modes at selected frequencies $f_i$ inside the bandgap. The zoom-in shows a conceptual view cutout of the resonant inclusions.

The model described by Eqns. (1) and (2) is used to illustrate the three main characteristics of the embedded resonant lens: (1) beamforming, (2) beamsteering, and (3) sub-wavelength resolution. These characteristics are shown using three separate examples.

In the first example, the plate was equipped with a semi-circular anisotropic resonant metamaterial lens with $n=6$ layers and a line defect located at $\theta=90°$. The line defect was created by altering the stiffness properties of the inclusions in a 20°-wide sector centered at $\theta$. The inclusions non-dimensional mass ratio was set to $m_r = m_{Resonator}/m_{Plate} = 6.\text{e-}3$ while the spacing to $s_r = 0.02$m and $s_\theta = s_r/2 = 0.01$m in the radial and azimuthal directions, respectively. The stiffness of the inclusions in the line defect was set to $0.2\times10^9$ N/m (from the initial $0.6\times10^9$ N/m) resulting in an uncoupled frequency of the individual resonator of $f_{90°}==5.032$kHz. The ultrasonic transducer was located in the center of the lens (red dot) and used to generate a 5.5 period Hanning-windowed tone burst (FIG. 4a, inset) tuned at the fundamental frequency f90° of the line defect. Numerical results, given in terms of squared wave amplitude distribution, show that the lens is able to convert the omni-directional wave produced by the single actuator into an ultrasonic beam (FIG. 4b) that remains collimated upon propagation in the plate. Without the lens the point excitation would produce a diffused wave field as shown in FIG. 4a.

The concept of line defect can also be exploited to steer the excitation beam (beamsteering). The lens can be spatially tailored in the azimuthal direction by embedding multiple line defects tuned at different frequencies. Each line defect is associated with well defined localized modes in the bandgap and can be activated by tuning the frequency of the excitation at the corresponding frequency of the defect. This concept is schematically illustrated in FIG. 3 where two sectors of the lens, centered at $\theta_1$ and $\theta_2$, are designed to host line defects associated with localized modes at frequencies $f_1$ and $f_2$. The previously developed thin plate model was used to numerically investigate this design. Two 20°-wide sectors centered at $\theta_1$=45° and $\theta_2$=135° and tuned at $f_1$=1.125kHz and $f_2$=7.957kHz were embedded in a circular lens located in the center of the plate. The inclusions non-dimensional mass ratio was set to $m_r$ =$m_{Resonator}$/$m_{Plate}$ = 6.e-4 while the spacing to $s_r$ = 0.04m and $s_\theta$ = $s_r$/2 = 0.01m. The lens was made of $n$=7 layers. The lumped stiffness of the inclusions in the two sectors was set to $k_1$= 1×10$^6$ N/m and $k_2$=5×10$^7$ N/m that resulted in fundamental uncoupled frequencies of the inclusions equal to $f_1$ and $f_2$, respectively.

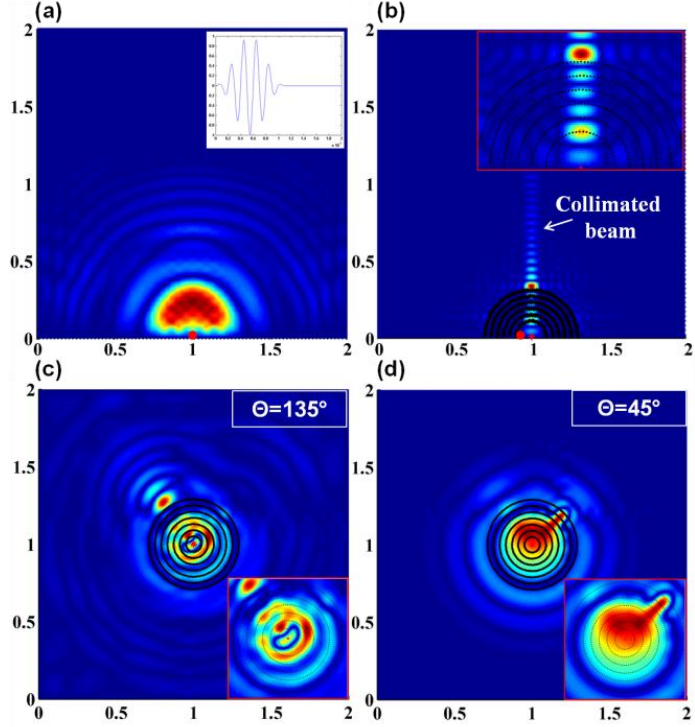

FIG. 4. Maps of the squared wave amplitude showing the performance of a spatially tailored resonant anisotropic metamaterial lens embedded in a thin simply supported aluminum plate. (b) shows the formation of the collimated beam due to a spatially tailored bandgap located at $\theta=90°$. The beam remains well collimated upon propagation in the host structure. Without the lens the source would produce a diffused wave field as in (a). (c) and (d) show that by exploiting the concept of spatially tailored bandgap the ultrasonic beam can also be steered by simply controlling the excitation frequency. The beam at $\theta_1=45°$ is triggered by an excitation at $f_1=1.125$kHz while the beam at $\theta_1=90°$ is triggered by an excitation at $f_2=7.957$kHz.

The tone burst excitation applied at the center of the lens was tuned first at $f_1$ and then at $f_2$. Numerical simulations (FIG. 4c, d) clearly show that the lens design is able to generate an ultrasonic beam in the predefined direction by simply tuning the frequency of the excitation. It is envisioned that the spatial tailoring of the bandgap could be implemented over the 360° span of the lens to achieve full control on the direction of the beam.

Embedded metamaterial lenses can also be used as a tool to increase the sensitivity during the damage sensing phase. In particular, anisotropic lenses can be used to achieve sub-wavelength resolution (beyond the diffraction limit) to clustered incipient damage. Incipient damage is known to induce nonlinear harmonic response[19] when excited by an intense ultrasonic field. Several studies have exploited this characteristic nonlinear response to perform remote damage identification and localization[20,21]. In case of clustered incipient damage, the narrow spacing between the nonlinear damage (sources) does not allow discerning their individual location from measurements in the far-field, where sensors are typically located. The reconstructed damage appears as an aggregate with an overestimated footprint and without meaningful information on the damage shape (useful for damage classification). This limited resolution is the result of a well known physical phenomenon in optics and acoustics: the *Abbe diffraction limit*[22]. In an acoustic (or optic) image the finer details (smaller than the characteristic wavelength at the frequency of interest) are associated with the evanescent waves generated upon diffraction of the incident wave from the scatterer. The diffracted evanescent waves decay exponentially in the near-field leading to a loss of information on the scatterer. It is this lost information, in fact, that limits the spatial resolution of sensors located in the far-field. Anisotropic metamaterials can convert evanescent into propagating waves and project them into the far-field[9-11]. This concept is extended to finite elastic structures in order to obtain sub-wavelength resolution of incipient damage. The performance of the lens is illustrated by using the embedded semi-circular design previously discussed. In this case no spatial tailoring is needed (uniform inclusions were used throughout the lens). The canalization effect[23] is exploited to convert evanescent into propagating waves. The lens is tested by using two incipient damage simulated by acoustic sources at a selected frequency $f_d$=5.032kHz. The two damage (sources) are separated by a distance $\varDelta s$=0.1m along the *x*-axis. This distance corresponds to a range of about $0.1\lambda \div \lambda$, where $\lambda$ is the wavelength of the $S_0$ and $A_0$ (Lamb) modes at the selected frequency. Results in FIG. 5b clearly show that the lens is able to project a distinct image of the two sources in the far-field therefore achieving a spatial resolution of at least $0.1\lambda$ that is beyond the diffraction limit.

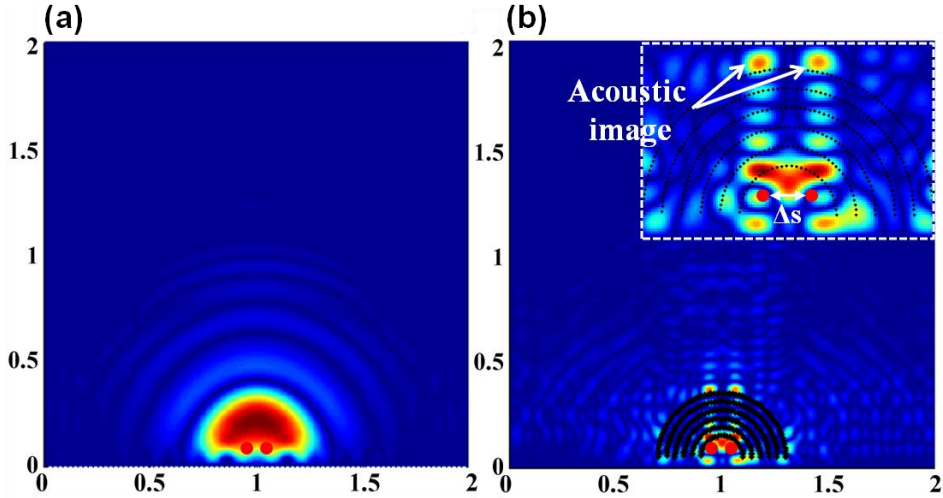

FIG. 5. Numerical results showing the sub-wavelength resolution capability in a thin plate with an embedded metamaterial lens. The two acoustic sources (red dots) are separated by a distance smaller than the diffraction limit. (a) without the lens the two sources generate a diffused field with no information about the individual sources. (b) with the lens, a distinct image of the original sources is projected to the far-field providing resolution beyond the diffraction limit.

The response of the plate without the lens (FIG. 5a) is also provided for comparison. In this case, the two sources produce a diffused field where the information on the individual sources is lost.

In conclusion, this Letter showed that the use of embedded resonant metamaterial lenses in structural elements can be used to achieve selective interrogation and high spatial resolution of clustered acoustic sources (such as those produced by nonlinear damage). The lens design exploits the characteristics of anisotropic resonant metamaterials and spatially tailored bandgaps.

Numerical results show that this approach is able to generate a steerable collimated (or focused) ultrasonic excitation by relying on the use of a single transducer.

[1] Semperlotti F., Wang K. W., Smith E. C., AIAA J. **47** (2009).

[2] Semperlotti F., Wang K.W., Smith E. C., Appl. Phys. Lett. **95,** 254101 (2009).

[3] Lamberti A, Semperlotti F., Smart. Mat. Struct, accepted for publication.

[4] Li, J., Rose, J. L., IEEE Trans. Ultras. Ferroel. Freq. Contr. **48** (2001).

[5] Yu, L., Giurgiutiu, V., Ultrasonics **48** (2008).

[6] Liu, Z., Zhang, X., Mao, Y., Zhu, Y. Y., Yang, Z., Chan, C. T., Sheng, P., Science **289** (2000).

[7] Li, J., Fok, L., Yin, X., Bartal, G., Zhang, X., Nature Materials **8** (2009).

[8] Ao, X., Chan, C. T., Phys. Rev. B 77, 025601 (2008).

[9] Salandrino, A., Engheta, N., Phys. Rev. B **74,** 075103 (2006).

[10] Zubin, J., Leonid, V. A., Narimanov, E., Optics Express **14** (2006).

[11] Zhang, X., Liu, Z., 2008, Nature Materials 7 (2008).

[12] Lai, Y., Zhang, X., Zhang, Z-Q., Appl. Phys. Lett. **79**, 3224 (2001).

[13] Mei, J., Zhengyou, L., Shi, J., Decheng, T., Phys. Rev. B **67**, 245107 (2003).

[14] Anderson, P. W., Phys. Rev. **109**, 1492 (1958).

[15] Robillard, J. F., Matar, O. B., Vasseur, J. O., Deymier, P. A., Stippinger, M., Hladky-Hennion, A. C., Pennec, Y., Djafari-Rouhani, B., Appl. Phys. Lett. **95**, 124104 (2009).

[16] Hsu, J. C., Wu, T. T., Phys. Rev. Lett. **74**, 144303 (2006).

[17] Zhu, H., Semperlotti F., AIP Advances **3**, 092121 (2013).

[18] Meirovitch, L., *Fundamentals of Vibrations*, McGraw-Hill, 2001.

[19] Solodov I Y, Ultrasonics **36** (1998).

[20] Solodov I Y, Wackerl J, Pfleiderer K, Busse G, Appl. Phys. Lett. **84** 5386 (2008).

[21] Kazakov V V, Sutin A, Johnson P A, Appl. Phys. Lett. **81** 646 (2002).

[22] Nettel S., *Wave Physics: Oscillation - Solitons - Chaos*, Springer 1995.

[23] He. Z., Cai F., Ding Y., Liu Z., Appl. Phys. Lett. **93**, 233503 (2008).